\documentclass[11pt]{article}
\usepackage[margin=0.9in]{geometry}

\usepackage{graphicx,url}

\usepackage[utf8]{inputenc}  

\usepackage{authblk} 

\usepackage{amsmath,amsfonts,amssymb}
\usepackage{natbib}
\usepackage{hyperref}
\usepackage{xcolor}
\usepackage{pdfpages}

\title{Benchmarking multi-step methods for the dynamic prediction of survival with numerous longitudinal predictors}

\author[1]{Mirko Signorelli}
\author[2]{Sophie Retif}

\affil[1]{Mathematical Institute, Leiden University (NL)}
\affil[2]{School of Industrial and Information Engineering, Politecnico di Milano (IT)}

\date{}

\begin{document} 

\maketitle

\subsection*{About this article}

\begin{itemize}
\item Please \textbf{cite this article as}: Signorelli, M., Retif, S. (2025). Benchmarking multi-step methods for the dynamic prediction of survival with numerous longitudinal predictors.  \textit{The International Journal of Biostatistics}. DOI: 10.1515/ijb-2025-0049.
\item This document contains the accepted version of the manuscript. The \textbf{final version} can be freely downloaded (Open Access) from the website of the IJB, using this link: \href{https://doi.org/10.1515/ijb-2025-0049}{https://doi.org/10.1515/ijb-2025-0049}.
\end{itemize}

\begin{abstract}
\noindent
In recent years, the growing availability of biomedical datasets featuring numerous longitudinal covariates has motivated the development of several multi-step methods for the dynamic prediction of survival outcomes. These methods employ either mixed-effects models or multivariate functional principal component analysis to model and summarize the longitudinal covariates' evolution over time. Then, they use Cox models or random survival forests to predict survival probabilities, using as covariates both baseline variables and the summaries of the longitudinal variables obtained in the previous modelling step.

Because these multi-step methods are still quite new,  to date little is known about their applicability,  limitations, and predictive performance when applied to real-world data. To gain a better understanding of these aspects, we performed a benchmarking of these multi-step methods (and two simpler prediction approaches) using three datasets that differ in sample size, number of longitudinal covariates and length of follow-up.  
We discuss the different modelling choices made by these methods, and some adjustments that one may need to do in order to be able to apply them to real-world data.  Furthermore, we compare their predictive performance using multiple performance measures and landmark times,  assess their computing time, and discuss their strengths and limitations.

\noindent \textbf{Keywords:} dynamic prediction; survival analysis; longitudinal data; risk prediction models; clinical prediction models.

\end{abstract}

\section{Introduction}
\label{sec1}

Predicting the probability that individuals may experience adverse events (such as the onset of a disease, or death) is an important task of modern medicine. Risk prediction models (RPMs, \citealp{steyerberg2009}) are statistical and machine learning models that can be employed to predict the probability that an individual (or patient) will experience an event of interest over time.  They can be used by clinicians to monitor disease progression,  provide patients with information about their health status and risks, and guide decisions about hospitalization, surgeries and initiating, adjusting or terminating a certain treatment.

The development of a RPM requires the estimation of the probability that individuals will not experience the event over time (called \textit{survival probability}) as a function of their risk factors (\textit{covariates} or \textit{predictors}).  Often,  all covariates are time-independent (i.e., they do not change over time) or measured at a single point in time.  The main advantages of this static modelling approach are that it is relatively easy to implement, and it does not require gathering patient information over time. However,  static RPMs do not incorporate any information on individual changes over time that may be predictive of the event of interest.
Dynamic RPMs make it possible to include longitudinal (``time-dependent'') covariates (i.e.,  covariates measured at multiple points in time on the same individual), empowering more accurate predictions in situations where changes in the level of one or more covariates over time are associated with the risk of experiencing the event. Moreover,  dynamic RPMs can be used to dynamically update individual predictions of survival each time new (more recent) measurements become available.

Traditional approaches to the dynamic prediction of survival outcomes include last observation carried forward (LOCF) landmarking \citep{vanhouwelingen2007} and joint modelling \citep{hickey2016}.  The major advantages of the LOFC approach are its simplicity and ease of implementation; possible disadvantages include the fact that the last observation taken before the landmark may be outdated, and the lack of a model to describe the evolution over time of the longitudinal covariates and to account for measurement error (which may be considerable in the case of biomarkers).  Unlike LOCF landmarking,  joint models (JMs) enable efficient use of the longitudinal data, and to account for measurement error.  
However,  the estimation of JMs is computationally difficult and demanding, particularly when attempting to include many longitudinal covariates in the JM.  
To date, high computing times and frequent convergence problems have limited the applicability of JMs to problems with a limited number of longitudinal covariates \citep{fournier2019,spreafico2021},  forcing researchers working on data with numerous longitudinal covariates to preselect a subset of those covariates before fitting the JM.

Over the last decades, technological advancements have led to a growing availability of studies where many longitudinal covariates are measured alongside a survival outcome. Examples include observational studies were patients are monitored for several years to study the occurrence of a slowly progressing disease such as cancer \citep{patel2017}, dementia \citep{bennett2018} or Alzheimer's disease \citep{weiner2010}, as well as studies that seek to identify biomarkers associated with disease progression and the occurrence of disease milestones among a large number of longitudinal omic variables \citep{signorelli2020,filbin2021}.
Recently, this increasing availability of repeated measurement data in biomedical studies has driven the development of new statistical methods \citep{li2019,signorelli2021, lin2021, devaux2023, gomon2024} and software \citep{signorelli2024, devaux2024} for dynamic prediction that can better handle the availability of a large number of longitudinal covariates.  These methods approach the dynamic prediction problem through a multi-step modelling  approach: first, they employ either mixed-effects models (MEMs) or multivariate functional principal component analysis (MFPCA) to describe the evolution over time of the longitudinal covariates and to obtain time-independent summaries of the repeated measurement data; then, they model the relationship between the survival outcome and the available covariates using either Cox models or random survival forests (RSFs).

Because these multi-step methods have been introduced very recently, to date little is known about their applicability to, and performance with, real-world datasets.  Practical modelling questions include how well these methods can accommodate real-world datasets with numerous longitudinal covariates and unbalanced measurement times, and how computationally intensive they are (especially when compared to a computationally inexpensive approach such as landmarking, and to a computationally inefficient one such as JM). From an applied perspective, it is also unknown what the predictive performance of the different multi-step methods may be when applied to real-world datasets.

The goal of this study is to investigate these open questions using publicly available real-world datasets.  We gathered data from three longitudinal studies that differ substantially in terms of sample size, number of longitudinal covariates and length of follow-up. 
We employed four multi-step dynamic prediction methods, alongside two simpler prediction approaches, to predict a relevant time-to-event outcome in each dataset, and proceeded to evaluate the predictive performance of each method according to three different metrics for survival data.  
Although an additional comparison with the JM approach would be interesting,  JMs are not included in this benchmarking study because their estimation through state of the art statistical software for JMs (such as \texttt{joineMRL} and \texttt{JMbayes2}) repeatedly failed when attempting to include more than three to five longitudinal predictors in the JM.

The remainder of this article is organized as follows: in Section \ref{sec2} we briefly describe the dynamic prediction problem and the methods considered in the benchmarking study.  In Section \ref{sec3} we present the datasets used for the benchmarking, and provide detailed information on how each prediction method was applied to each dataset. Section \ref{sec4} presents the results of the benchmarking,  which are further discussed in Section \ref{sec5}.

\section{Methods}
\label{sec2}

\subsection{Dynamic prediction of survival outcomes}
\label{ss:notation}

We consider the setup of a longitudinal study where $n$ subjects are enrolled, and interest lies in predicting a time-to-event outcome. 
Each subject $i \in \{1, 2, ..., n\}$ is followed from time $t = 0$ (study entry / baseline visit) until they either experience the event of interest at time $t_i$, or they are right-censored at time $c_i$.
We denote by $T_i$ and $C_i$ the random variables corresponding to the event and censoring times,  respectively, and let $T_i^*=\min{\left(T_i, C_i\right)}$.
Furthermore, we denote by $t_i^*$ the observed value of $T_i^*$, and introduce the censoring indicator $\delta_i$ that is $1$ if for subject $i$ the event of interest is observed at $t_i^* = t_i$, and $0$ in case of censoring, i.e., if the event has not yet been observed at $t_i^* = c_i$.

Between times 0 and $t_i^*$,  subject $i$ undergoes $m_i \geq 1$ visits at times $t_{i1},  t_{i2}, ..., t_{i m_i}$, where we assume that $t_{i1} = 0$,  $t_{i1}<t_{i2}<\ldots<t_{im_i}$ and $t_{im_i} < t_i^*$.
At time $t_{i1}=0$, $P$ baseline covariates $x_i = (x_{1i}, ..., x_{Pi})^T$ are measured. Moreover, $Q$ longitudinal covariates are measured at each follow-up visit (including $t_{i1}=0$). We denote by $y_{qij}=y_{qi}(t_{ij})$ the value of the $q$-th longitudinal covariate measured on subject $i$ at time $t_{ij}$, and let $y_{ij} = y_i(t_{ij}) = (y_{1ij}, ..., y_{Qij})^T$ be the corresponding Q-dimensional vector.

Let $\mathcal{I}(t) = \{i \in \{1, 2,..., n\}: t_i^* > t\}$ denote the set of individuals that are still at risk at a given time point $t$.
Moreover, let $\mathcal{H}_i(t) = \left( x_i,  y_i(0),  ...,  y_i(t_{ik}) \right)$, where $k$ is such that $t_{ik} \leq t < t_{i, k+1}$,  denote the covariate information available for a subject $i \in \mathcal{I}(t)$ up until time $t$ (in other words, $\mathcal{H}_i(t)$ contains the $P$ baseline covariates and all repeated measurements taken up until time $t$ of the $Q$ longitudinal covariates).

At any point in time $t_1 \in (0, t_i^*)$, we may want to predict the conditional probability $P(T_i > t_2 | T_i > t_1)$ that individual $i \in \mathcal{I}(t_1)$ will still be event-free at time $t_2 > t_1$, given that they had not experienced the event at $t_1$.  More specifically,  given a so-called \textit{landmark time} $\ell$, our goal is to estimate the conditional survival function

\begin{equation}
S_i(t | \ell,  \mathcal{H}_i(\ell)) =  P(T_i > t | T_i > \ell, \mathcal{H}_i(\ell)),  \: i \in \mathcal{I}(\ell), \: t \geq \ell,
\label{eq:cond_surv}
\end{equation}

using all the baseline and longitudinal covariate information $\mathcal{H}_i(\ell)$ available up until $\ell$  for subjects $i \in \mathcal{I}(\ell)$. 

When developing dynamic RPMs,  it can often be of interest to compute predictions of survival $S_i(t | \ell_k,  \mathcal{H}_i(\ell_k))$ over a range of $K$ landmark times $\ell_1, \ell_2, ..., \ell_K$, rather than for a single landmark $\ell$.  In such situations, our notation can be easily expanded by defining $\mathcal{I}(\ell_k)$ and $\mathcal{H}_i(\ell_k)$ separately for each landmark time $\ell_k$.

\subsection{Strict and relaxed data landmarking}

Equation \eqref{eq:cond_surv} states that the goal of dynamic prediction is to predict survival probabilities over time for all individuals who are event-free up until the landmark time $\ell$, using as predictors all covariate values measured up to time $\ell$.

In practice, often repeated measurement data gathered after $\ell$ are available when estimating a statistical model, raising the question of whether such data should be used for model estimation.  \citet{gomon2024} named ``relaxed landmarking" the approach where longitudinal data gathered after the landmark are used to estimate the model on the training set, contrasting it to a ``strict landmarking'' method which discards any repeated measurement taken after $\ell$ in line with equation \eqref{eq:cond_surv}. They showed that using relaxed data landmarking led to a considerable worsening of the predictive performance of MFPCCox, when compared to the same model estimated using strict data landmarking.

At first sight, relaxed data landmarking may appear attractive, because it allows to use more data to estimate the model. However,  this approach is problematic for two reasons: first, it uses future information (collected after $t > \ell$) to predict the conditional survival probability $P(T > t | T > \ell)$.  Note that such future information is not available at the landmark time, and thus in reality it would not be possible to make use of it to make predictions at $\ell$. Furthermore, relaxed data landmarking introduces a selection bias in the modelling of the longitudinal trajectories: because individuals who survive longer are likely to have more repeated measurements taken after $\ell$ than individuals with shorter survival, the available measurements after the landmark are not a representative sample from the population of individuals who survived up until $\ell$.  Whereas this is not an issue with joint models, the multi-step approaches considered in this article do not correct for this source of bias. 

Despite these issues, in the literature relaxed data landmarking is often used to estimate multi-step dynamic prediction models. Based on the aforementioned remarks and the results of \citet{gomon2024},  hereafter we describe and implement the dynamic prediction models included in our comparison following a strict data landmarking approach.

\subsection{Dynamic prediction methods}

\subsubsection{Static Cox model}

Before describing the dynamic prediction methods compared in this work, we consider a static Cox model \citep{cox1972} that only uses baseline information to predict $S_i(t | \ell,  \mathcal{H}_i(\ell))$:

\begin{equation}
h_i(t) = h_0(t) \exp \left\{\alpha^T x_i + \gamma^T y_i(0) \right\}, \: i \in \mathcal{I}(\ell),
\label{eq:staticCox}
\end{equation}

where $h_i(t)$ denotes the hazard for subject $i$ at time $t$,  $h_0(t)$ is a non-parametric baseline hazard, and $\alpha$ and $\gamma$ are vectors  of regression coefficients. 
The predicted conditional survival probability of equation \eqref{eq:cond_surv} can be estimated by computing 
\begin{equation}
\hat{S}_i(t | \ell, \mathcal{H}_i(\ell) ) = \exp \left\{ - \int_0^t \hat{h}_0(v) e^{\hat{\eta}_i} dv  \right\},
\label{eq:predCox}
\end{equation}
where $\hat{\eta}_i$ is the estimated linear predictor for subject $i$ from model \eqref{eq:staticCox}.

Admittedly, model \eqref{eq:staticCox} does not make use of the available longitudinal measurements to update predictions of survival: nevertheless, it is useful to evaluate the predictive performance of this model to quantify how accurate our predictions of survival would be if we were to ignore any information collected after baseline. By comparing the performance of this static prediction model to that of dynamic methods, which incorporate more and more repeated measurements as the landmark time increases, we can effectively quantify the extent to which adding longitudinal information to a RPM may improve predictive performance. Although the discrepancy between static and dynamic methods will be typically data-dependent,  in general we can expect the difference between these two approaches to increase as the landmark time increases,  as the information provided to model \eqref{eq:staticCox} becomes more and more outdated, while dynamic methods continue being fed more and more up to date patient information.

\subsubsection{Landmarking}

Landmarking \citep{vanhouwelingen2007} is a pragmatic approach to dynamic prediction that summarizes the trajectory of each longitudinal covariate up until the landmark time into a single summary measure per covariate. Two commonly-used summary measures are the last available observation (up to $\ell$) of the longitudinal covariate, an approach referred to as \textit{last observation carried forward} (LOCF), and the average of all repeated measurements taken up until $\ell$. The summaries of the longitudinal covariates are then included as predictors in a Cox model alongside the baseline covariates.

Let $y_{qi} = (y_{qi1}, y_{qi2}, ..., y_{qim_i})^T$ denote all repeated measurements of the $q$-th longitudinal covariate for subject $i \in \mathcal{I}(\ell)$,  and denote by $y_{qi}^{\ell}$ the last non-missing value in $y_{qi}$ such that $t_{ik} \leq \ell$.
The LOCF landmarking method proceeds to fit the following Cox model
\begin{equation}
h_i(t) = h_0(t) \exp \left\{\alpha^T x_i + \sum_{q=1}^Q \gamma_q y_{qi}^{\ell} \right\}, \: i \in \mathcal{I}(\ell),
\label{eq:LOCF}
\end{equation}
where $\alpha$ and $\gamma = (\gamma_1, \gamma_2, ..., \gamma_Q)^T$ are vectors of regression coefficients.
The predicted survival for subject $i$ can be estimated using equation \eqref{eq:predCox}, where $\hat{\eta}_i$ is now the estimated linear predictor from model \eqref{eq:LOCF}.

\subsubsection{Multivariate Functional Principal Component Analysis Cox model (MFPCCox)}
\label{sss:mfpccox}

The Multivariate Functional Principal Component Analysis Cox (MFPCCox) model \citep{li2019,gomon2024} uses Multivariate Functional Principal Component Analysis (MFPCA) to summarize the $Q$ longitudinal covariates into MFPCA scores, which are then used as predictors of survival in a Cox model.

First, MFPCA \citep{happ2018} is employed to approximate the longitudinal trajectory for the $q$-th longitudinal predictor as the sum of a mean process $\mu_q(t)$ shared by all subjects, and a finite sum of subject-specific MFPCA scores $\rho_{ki}$ that are shared across the $Q$ longitudinal covariates:

\begin{equation}
y_{qij} =  y_{qi} (t_{ij}) \approx \mu_q (t_{ij}) + \sum_{k=1}^{K} \rho_{ki} \psi_{kq} (t_{ij}), \: i \in \mathcal{I}(\ell), \: t_{ij} \leq \ell,
\label{eq:MFPCA}
\end{equation}

where $\psi_{k} (t) = \left( \psi_{k1} (t), ..., \psi_{kQ} (t) \right)^T$ are $Q$-variate orthonormal eigenfunctions with associated MFPCA scores $\rho_{ki}$. The eigenfunctions are ordered by decreasing percentage of variance explained, and $K$ is chosen in such a way that the first $K$ eigenfunctions explain a certain proportion of the variance of the longitudinal covariates.

This is achieved by assuming that the observed $Y_{q}(t)$ is a noisy realization from an underlying random process $R_{q}(t)$ with mean $\mu_q (t) = E[R_{q}(t)] $ and covariance function $C_q (t, t') = Cov(R_q(t), R_q(t'))$: $Y_{q}(t) = R_{q}(t) + \epsilon_{q}(t)$, where $\epsilon_{q}(t)$ are i.i.d.  gaussian errors. 
First, one needs to estimate the mean $\mu_q (t)$ and covariance $C_q (t, t')$ functions using non-parametric estimators. Then,  $C_q (t, t')$ is decomposed to obtain the eigenfunctions, and the corresponding univariate FPCA scores are computed using the Principal component Analysis by Conditional Expectation (PACE) algorithm \citep{yao2005}. Lastly,  estimates of the multivariate FPCA scores $\rho_{ki}$ and eigenfunctions $\psi_{kq}(t)$ in equation \eqref{eq:MFPCA} are computed from the univariate FPCA scores \citep{happ2018}.

To predict survival,  MFPCCox considers a Cox model where the estimated MFPCA scores are included in the linear predictor alongside the baseline covariates:

\begin{equation}
h_i(t) = h_0(t) \exp \left\{\alpha^T x_i + \sum_{k=1}^{K} \gamma_k \hat{\rho}_{ki} \right\}, \: i \in \mathcal{I}(\ell),
\label{eq:MFPCCox}
\end{equation}

where $\hat{\rho}_{ki}$ are the estimated MFPCA scores, and $\alpha$ and $\gamma = (\gamma_1, \gamma_2, ..., \gamma_K)^T$ are vectors of regression coefficients.  
The predicted survival for subject $i$ can be estimated using equation \eqref{eq:predCox}, where $\hat{\eta}_i$ is now the estimated linear predictor from model \eqref{eq:MFPCCox}.

\subsubsection{Penalized Regression Calibration (PRC)}
\label{sss:prc}

The Penalized Regression Calibration (PRC) method \citep{signorelli2021,signorelli2024} uses mixed-effects models to summarize the $Q$ longitudinal covariates into predicted random effects, which are then used as predictors of survival in a Cox model.

First, PRC models each longitudinal covariate (or a transformed version of it, for example its log-transform) using a linear mixed effects model (LMM, \citealt{mcculloch2004}):

\begin{equation}
y_{qij} =  y_{qi} (t_{ij}) = W_{qi} (t_{ij}) \beta_q + Z_{qi} (t_{ij}) u_{qi} + \varepsilon_{qi}, \: i \in \mathcal{I}(\ell),  \: t_{ij} \leq \ell,
\label{eq:lmm}
\end{equation}

where $\beta_q$ is a vector of population parameters shared across all subjects called \textit{fixed effects},  $u_{qi}$ is a vector of subject-specific parameters called \textit{random effects} that are assumed to follow a multivariate normal distribution,  $\varepsilon_{qi}$ is a vector of Gaussian errors, and $W_{qi}(t_{ij})$ and $Z_{qi}(t_{ij})$ are design matrices containing the covariates relevant for $\beta_q$ and $u_{qi}$, respectively.  Notice that \citet{signorelli2021} also considered a more complex multivariate mixed model that is not relevant for the datasets considered in this paper.

Specifically, in this article we consider an LMM with a random intercept $u_{qi0}$ and a random slope $u_{qi1}$ associated to the time variable $t_{ij}$:

\begin{equation}
y_{qij} = \beta_{q0} + u_{qi0} + \beta_{q1} t_{ij} + u_{qi1} t_{ij} + \varepsilon_{qij}, \: i \in \mathcal{I}(\ell),  \: t_{ij} \leq \ell,
\label{eq:lmmwithslope}
\end{equation}

where $\beta_q = (\beta_{q0}, \beta_{q1})^T$,  $u_{qi} = (u_{qi0}, u_{qi1})^T \sim N(0, \Sigma_q)$, and $\varepsilon_{qi} \sim N(0, \sigma^2_q I)$. Notice that although \citet{signorelli2021} and \citet{signorelli2024} suggest to employ age at observation as time covariate in equation \eqref{eq:lmmwithslope},  here we choose to model $y_{qij}$ as a function of time on study $t_{ij}$ to make PRC directly comparable with MFPCCox, FunRSF and DynForest, as all of those methods model $y_{qij}$ on the time on study scale. 

After model \eqref{eq:lmm} has been estimated,  PRC proceeds to compute the vector with the predicted random effects  $\hat{u}_{qi}$ for each covariate. As an example,  for model \eqref{eq:lmmwithslope} PRC computes a predicted random intercept $\hat{u}_{qi0}$ and slope $\hat{u}_{qi1}$, which are subsequently used as summaries of the vector $y_{qi}$.

To predict survival,  PRC uses a Cox model where the predicted random effects are used as predictors of survival together with the baseline covariates:
\begin{equation}
h_i(t) = h_0(t) \exp \left\{\alpha^T x_i + \sum_{q=1}^{Q} \gamma_q \hat{u}_{qi0} + \sum_{q=1}^{Q} \theta_q \hat{u}_{qi1}  \right\}, \: i \in \mathcal{I}(\ell),
\label{eq:PRC}
\end{equation}
where $\hat{u}_{qi0}$ and $\hat{u}_{qi1}$ denote the estimated random effects from model \eqref{eq:lmmwithslope},  and $\alpha$, $\gamma = (\gamma_1, ..., \gamma_Q)^T$ and $\theta = (\theta_1, ..., \theta_Q)^T$ are vectors of regression coefficients.
Model \eqref{eq:PRC} is estimated using penalized maximum likelihood.
The predicted survival for subject $i$ can be estimated using equation \eqref{eq:predCox}, where now $\hat{\eta}_i$ denotes the estimated linear predictor from model \eqref{eq:PRC}.

\subsubsection{Functional Random Survival Forest (FunRSF)}
\label{sss:funrsf}

The Functional Random Survival Forest (FunRSF) approach \citep{lin2021} uses MFPCA to summarize the trajectories described by the longitudinal covariates, and a random survival forest (RSF, \citealp{iswaran2008}) to predict survival.

The MFPCA modelling step is analogous to the one used by MFPCCox and described in Section \ref{sss:mfpccox}. This step yields $K$ estimated MFPCA scores $\hat{\rho}_{ki}, k = 1, ..., K$ for each subject $i \in \mathcal{I}(\ell)$, which are then used within the RSF as time-independent covariates together with the baseline covariates $x_i$.

The estimation of the RSF involves sampling $B$ bootstrap samples from a dataset that comprises information on the survival outcome, the baseline covariates and the MFPCA scores. For each boostrap sample, a random survival tree is estimated using as splitting variables both the baseline covariates, and the MFPCA scores. 
At each node of the tree, a fixed number $F$ of candidate splitting variables is selected among the union of the $P$ baseline covariates and $K$ MFPCA scores. Each node is split by identifying the candidate variable whose splitting maximizes the log-rank test statistic, and the splitting is continued until a terminal node is reached.  A node is terminal if further splitting would lead to child nodes that contain less than a fixed number $s$ of subjects. 
Each terminal node $k$ in the $b$-th tree is then summarized by its cumulative hazard function (CHF) $\hat{H}_{bk}(t | \ell)$. The estimated CHF for subject $i$ is obtained by dropping $i$ through each tree to obtain its estimated CHF $\hat{H}_b(t | \ell,  x_i, \hat{\rho}_i )$, and then averaging over the $B$ bootstrap samples to obtain the ensemble CHF estimate

\begin{equation}
\hat{H}(t |\ell, x_i, \hat{\rho}_i) ) = \frac{1}{B} \sum_{b = 1}^B \hat{H}_b(t | \ell, x_i, \hat{\rho}_i),
\label{eq:ensembleCHF-FunRSF}
\end{equation}

which is then used to estimate $\hat{S}_i \left( t | \ell, \mathcal{H}_i(\ell) \right) = \exp \left\{ \hat{H}(t |\ell, x_i, \hat{\rho}_i) \right\}$.

\subsubsection{Dynamic Random Survival Forest (DynForest)}
\label{sss:dynforest}

The Dynamic Random Survival Forest (DynForest) method  \citep{devaux2023,devaux2024} makes use of LMMs to summarize the $Q$ longitudinal covariates into predicted random effects that are used as predictors of survival within an RSF.

The estimation of the RSF starts by drawing B bootstrap samples from a dataset that comprises information on the survival outcome, the baseline covariates and the longitudinal covariates. For each bootstrap sample, a survival tree is estimated as follows: at each node,  a fixed number $F$ of candidate splitting variables is selected among the union of the $P$ baseline covariates and the $Q$ longitudinal covariates. 
For each candidate longitudinal covariate, an LMM of the form given in equation \eqref{eq:lmm} is estimated,  and the longitudinal trajectory of covariate $q$ for subject $i$, $y_{qi}$, is then summarized by the predicted random effects $\hat{u}_{qi}$ in the same way as done by PRC (see Section \ref{sss:prc}). 
The predicted random effects from the candidate longitudinal covariates are then used as time-independent covariates alongside the candidate baseline covariates as possible splitting variables for the given node.  Similarly to FunRSF, the node is split choosing the candidate covariate that maximizes the log-rank statistics.
The construction of the tree continues until a terminal node that contains a pre-specified number of subjects $s$, and a prespecified number of events $e$ is reached. Lastly, the estimated CHF for subject $i$ is computed by dropping the individual down each tree, obtaining their estimated CHF according to the $b$-th tree, $\hat{H}_{b}(t | \ell, x_i,  \hat{u}_i)$,  and averaging those tree-specific CHFs to obtain the ensemble CHF estimate

\begin{equation}
\hat{H}(t |\ell, x_i,  \hat{u}_i) = \frac{1}{B} \sum_{b = 1}^B \hat{H}_b(t | \ell, x_i,  \hat{u}_i),
\label{eq:ensembleCHF-DynForest}
\end{equation}

which is then used to estimate $\hat{S}_i \left( t | \ell, \mathcal{H}_i(\ell) \right) = \exp \left\{ \hat{H}(t |\ell, x_i,  \hat{u}_i) \right\}$.

Notice that although the construction of the RSF in DynForest shares similarities with FunRSF,  an important difference is that FunRSF uses a single MFPCA computation to summarize the longitudinal covariates into MFPCA scores prior to estimating the RSF, whereas DynForest estimates new LMMs within each tree and node (meaning that the same longitudinal covariate may be summarized by different predicted random effects across trees and nodes). 
Differently from MFPCCox, PRC and FunRSF, DynForest can handle competing risks (which we do not address this in this article,  since most multi-step methods have not yet been extended to the competing risks setting).

\subsubsection{Software implementation}

All computations presented in this study were performed using \texttt{R} on an AMD EPYC 7662 processor with 2 GHz CPU.  
After properly subsetting the data as appropriate for the static Cox model and LOCF landmarking model, we proceeded to implement such models in \texttt{R}, relying on the \texttt{coxph} function from \texttt{R} package \texttt{survival} \citep{therneau2000} for model estimation.
We used on the \texttt{R} packages \texttt{pencal} \citep{signorelli2024} and \texttt{DynForest} \citep{devaux2024} to implement and estimate the PRC and DynForest models.
Lastly,  to estimate MFPCCox and FunRSF we proceeded to adapt as appropriate the code provided in \citet{li2019} and \citet{lin2021}, respectively.
All scripts created to perform the analyses presented in this paper are publicly available at \url{https://github.com/mirkosignorelli/comparisonDynamicPred}.

\section{Datasets and implementation}
\label{sec3}

Sections \ref{ss:ADNIdata}-\ref{ss:PBC2data} present the three datasets used for the benchmarking.  We describe how the data were gathered, the length of follow-up and the frequency of repeated measurements,  define the survival outcome that we want to predict, and list the baseline and longitudinal covariates that are employed to predict it.
The datasets differ substantially in  sample size, length of the follow-up, frequency and number of repeated measurements per subject, and number of longitudinal covariates employed as predictors (Supplementary Table 1),  making them representative of different scenarios that one may encounter when dealing with real-world longitudinal studies.

Lastly, in Section \ref{ss:implementation} we provide information about the implementation of the dynamic prediction methods for these datasets, including information about landmark times and prediction horizons, performance measures, and modelling and computational details.

\subsection{The ADNI dataset}
\label{ss:ADNIdata}

The Alzheimer's Disease Neuroimaging Initiative (ADNI) study \citep{weiner2010} is an ongoing study started in 2004 that was designed to identify and validate biomarkers related to the progression of AD.  By April 2023, the study had enrolled 2428 individuals, each scheduled to undergo an initial visit and subsequent assessments scheduled at 3, 6, 12, 18 and 24 months from baseline, followed by annual visits thereafter. At each follow-up visit, participants undergo comprehensive evaluations for dementia, including a variety of cognitive assessments, biospecimen sampling and brain imaging analysis.

Differently from the ROSMAP study that will be introduced in Section \ref{ss:ROSMAPdata},  where the cause of dementia (AD or not AD) is reported,  the ADNI study only reports dementia diagnoses, without attributing them to AD or other causes.  At each visit, ADNI participants are classified into one of the following categories: cognitive normal (CN),  mildly cognitive impaired (MCI), or affected by dementia. Moreover, both baseline information about the participant and longitudinal measurements of cognitive, imaging and biochemical markers are collected.

Our goal is to predict the time until a dementia diagnosis for subjects that entered the study as CN or MCI. To predict this survival outcome we employ 5 baseline covariates (age, gender, baseline diagnosis, number of apolipoprotein $\epsilon$4 alleles, and number of years of education) and 21 longitudinal covariates that are listed in Supplementary Table 2 (these covariates were selected based on their relevance as predictors of dementia, provided that they did not have too many missing values).
After removing subjects already diagnosed with dementia at baseline, subjects without any follow-up information after baseline, and subjects with missing covariate values at the baseline visit, 1643 subjects were retained for analysis.  
The number of visits per subject varies from 1 to 22,  with a mean of 6.2 visits. The follow-up length ranges from 3 months to 15.5 years, with a mean of 3.4 years.
Supplementary Figure 1 illustrates how the number of individuals at risk evolves over time, and the Kaplan-Meier estimator of the probability to be free from dementia in this dataset.

\subsection{The ROSMAP dataset}
\label{ss:ROSMAPdata}

The Religious Orders Study and Rush Memory and Aging Project (ROSMAP) study \citep{bennett2018} is an ongoing project designed to study the onset of Alzheimer's disease (AD).  Started in 1994,  by May 2023 ROSMAP had enrolled 3757 participants, collecting longitudinal information about their cognitive status and possible risk factors for AD.

Every year, subjects enrolled in ROSMAP undergo a clinical assessment that leads to an evaluation of their cognitive status, which is classified into 6 classes: no cognitive impairment (NCI); mild cognitive impairment (MCI); MCI and another condition contributing to cognitive impairment (MCI+); Alzheimer's dementia (AD); AD and other condition contributing to CI (AD+); other primary cause of dementia,  without clinical evidence of AD (Other).  In addition,  longitudinal information about a wide range of genomic, experiential, psychological and medical risk factors is collected alongside information  about the participant's background (such as age, gender and education).

We aim to predict the time until AD is diagnosed (i.e., cognitive status AD or AD+) for subjects that entered the study in the NCI, MCI, MCI+ and Other categories. To do so, we employ 5 baseline covariates (age, gender, years of education received, baseline cognitive status and presence of a cancer diagnosis) and a set of 30 longitudinal covariates, listed in Supplementary Table 3, which measure aspects related to several different domains (these covariates were selected from a wider list of longitudinal covariates based on their relevance as risk factors for AD,  provided that they did not have too many missing values).
After removing subjects already diagnosed with AD at baseline, subjects without any follow-up information after baseline, and subjects with missing covariate values at baseline, 3293 subjects were retained for analysis.  
The number of visits per subject varies from 2 to 30,  with a mean of 9.3 visits. The follow-up period ranges from 1 to 29 years, with a mean of 8.3 years. 
Supplementary Figure 2 illustrates how the number of individuals at risk evolves over time, and the Kaplan-Meier estimator of the probability to be free of AD in the ROSMAP dataset.

\subsection{The PBC2 dataset}
\label{ss:PBC2data}

The PBC2 dataset contains data from a clinical trial on primary biliary cholangitis (PBC) conducted between 1974 and 1984 at the Mayo Clinic \citep{murtaugh1994}. The randomized trial involved 312 participants who were randomized between a placebo group, and a treatment group where patients received the drug D-penicillanime. 
The trial recorded the occurrence of the first of two survival outcomes: liver transplantation, and death.  Because most of the multi-step methods considered in our comparison cannot deal with competing risks, for the purpose of our analysis we focus on predicting time to death, treating patients who underwent liver transplantation as right-censored at the date of transplantation.

To predict time to death, we employ 3 baseline covariates (age, gender and treatment group) alongside the 8 longitudinal covariates listed in Supplementary Table 4. Patients enrolled in the trial underwent visits upon study entry,  6 and 12 months after baseline, and yearly visits after that. 
The number of follow-up visits per patient ranged from 1 to 22, with an average of 6.1; the follow-up period ranged from 0.1 to 14.1 years, with a mean of 4.6 years.
Supplementary Figure 3 shows the number of patients at risk and the Kaplan-Meier estimator of the survival probability for the PBC2 dataset.

\subsection{Application of the dynamic prediction methods to the ADNI, ROSMAP and PBC2 datasets}
\label{ss:implementation}

\subsubsection{Definition of the landmark and horizon times}

The datasets considered in our study differ substantially in their sample size,  frequency of the repeated measurements (visits) and length of the follow-up.  All these aspects have an influence on the number of individuals still at risk at a given landmark, and the number of repeated measurements per subject available up to the landmark time.  Lastly, the length of the follow-up across patients affects the possibility to evaluate predictive performance at a given horizon time.

Given all these differences, we determined the landmark and horizon times at which predictions of survival are evaluated separately for each dataset,  striving to balance the need to have a good number of repeated measurements to model the evolution of the longitudinal covariates with MFPCA or LMMs, and that of having a good number of individuals still at risk to be able to estimate the Cox / RSF model and reliably assess predictive performance.  Clearly, this is easier for the ADNI and ROSMAP datasets, whose sample sizes are considerably bigger than that of PBC2.

The landmarks considered in our analysis are the following:

\begin{enumerate}
\item ADNI: 2, 3 and 4 years from baseline;
\item ROSMAP: 2, 3, 4, 5 and 6 years from baseline;
\item PBC2: 2.5, 3 and 3.5 years from baseline.
\end{enumerate}

The accuracy of predictions was evaluated for every year after the landmark up to 10 years from baseline for ADNI,  up to 15 years from baseline for ROSMAP,  and up to 8 years from baseline for PBC2.

\subsubsection{Performance measures and validation of predictive performance}

To evaluate predictive performance, we consider the three most commonly-used measures of predictiveness in survival analysis, namely the Brier score \citep{graf1999}, the time-dependent AUC \citep{heagerty2000}, and the C index \citep{pencina2004}.

We employed \textit{repeated} cross-validation (RCV) to obtain unbiased estimates of the predictive performance of the different models. For the ROSMAP and ADNI datasets we employed 10-fold cross-validation (CV) with 10 repetitions; for the PBC2 dataset, where using 10-fold CV led to very small validation folds, we employed 5-fold CV with 20 repetitions.

\subsubsection{Implementation details}

To ensure that every method was trained on data from the same set of subjects,  in each dataset we excluded subjects for which (a) the value of one or more baseline covariates was missing or (b) the value of one of the longitudinal predictors at the baseline visit was missing.
Condition (a) is needed by all methods to ensure that there are no missing values in the baseline covariates,  whereas (b) aims to ensure that are no missing values when estimating the static Cox model (where we use the measurements of the longitudinal covariates at the baseline visit as baseline covariates).
Besides this data selection step,  the four multi-step methods required some additional preprocessing and modelling choices that are specific to each method.

MFPCCox and FunRSF rely on MFPCA to summarize the longitudinal covariates into MFPCA scores.  In principle, MFPCA can handle longitudinal data where the measurement times differ across subjects. However,  highly-irregular measurement times complicate the estimation and inversion of the univariate covariance functions $C_q(t, t')$, often leading to numeric errors when attempting to perform the MFPCA step as implemented in MFPCCox and FunRSF.
Given that visits in the ADNI and ROSMAP datasets follow a regular measurement grid, but in a few instances the actual visit takes place a few days before or after the planned date, we proceeded to align the measurement times across subjects so as to obtain a regular measurement grid that allowed to solve most of the problems with the estimation of the MFPCA scores.
Furthermore, the MFPCA step within MFPCCox and FunSRF requires the specification of the minimum percentage of variance explained (PVE) to retain when reducing the longitudinal covariates to univariate FPCA scores (PVE$_1$), and then from univariate to multivariate FPCA scores (PVE$_2$). We proceeded to set PVE$_1 =$ PVE$_2 = 90\%$ for both methods.

DynForest and PRC employ LMMs to summarize the longitudinal covariates into predicted random effects.  When the response variable is strongly skewed, this can often lead to estimation (convergence) problems.  To reduce the frequency of such convergence problems and the need to address convergence issues with ad-hoc solutions each time they arise during the RCV, we proceeded to apply a log-transformation to those longitudinal covariates whose skewness index was larger than 1, and a cubic transformation to those with skewness index below -1.

PRC allows to choose between the ridge, elasticnet and lasso penalty for the estimation of the Cox model in equation \eqref{eq:PRC}. In this study we used the ridge penalty, which is the default option in the \texttt{pencal} package and the recommended choice in \citet{signorelli2021}.

Lastly, the RSF algorithms used by FunRSF and DynForest require the specification of multiple parameters. 
As concerns FunRSF,  the random forests were trained using an ensemble of $B = 1000$ trees.  The minimum number of subject in a terminal node $s$ was required to be at least 15 for all datasets. The number of predictors $F$ to be randomly selected at each node was set to the square root of the total number of predictors available in the given dataset.
The random forests within DynForest were trained using $B  = 200$ trees because estimating DynForest is much more time-consuming than estimating FunSRF (see Section \ref{res:time}).  Choosing the $F$, $s$ and $e$ parameters for this method is more complicated than for FunRSF, because the need to re-estimate the LMMs after each node split often leads to convergence problems within DynForest.  To reduce the frequency of convergence problems during the RCV, multiple values of $s$ (listed in Supplementary Table 5) were used sequentially in case of convergence problems,  progressively increasing $s$ to reduce the chance of encountering convergence errors towards the end of the tree.  The value of $F$ was set in the same way as for FunRSF; $e$ was set equal to 5 for ROSMAP and ADNI, and to 4 for PBC2.

\section{Results}
\label{sec4}

We now present the results of our benchmarking study. In Sections \ref{res:ADNI}, \ref{res:ROSMAP} and \ref{res:PBC2} we compare the predictive performance of the multi-step dynamic prediction methods on the ADNI, ROSMAP and PBC2 datasets, respectively.  In Section \ref{res:time}, we turn our attention to computing time.

\subsection{Predictive performance on the ADNI data}
\label{res:ADNI}

\begin{figure}[h!]
\begin{center}
\includegraphics[scale=0.8]{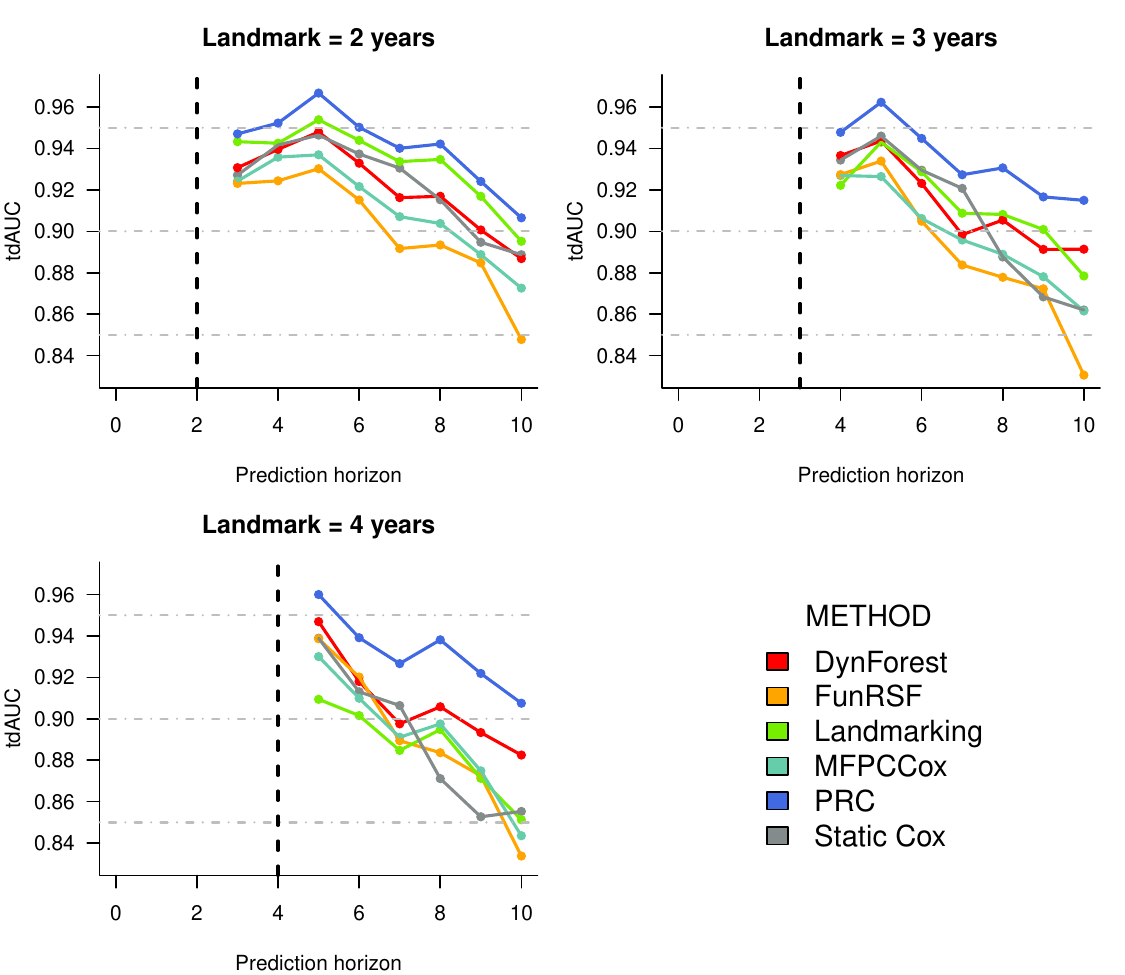}
\end{center}
\caption{Cross-validation estimates of the time-dependent AUC for the prediction of time to dementia in the ADNI dataset. 
The corresponding numeric values can be found in Supplementary Table 6.}
\label{ADNI-AUC}
\end{figure}

\begin{figure}[h!]
\begin{center}
\includegraphics[scale=0.8]{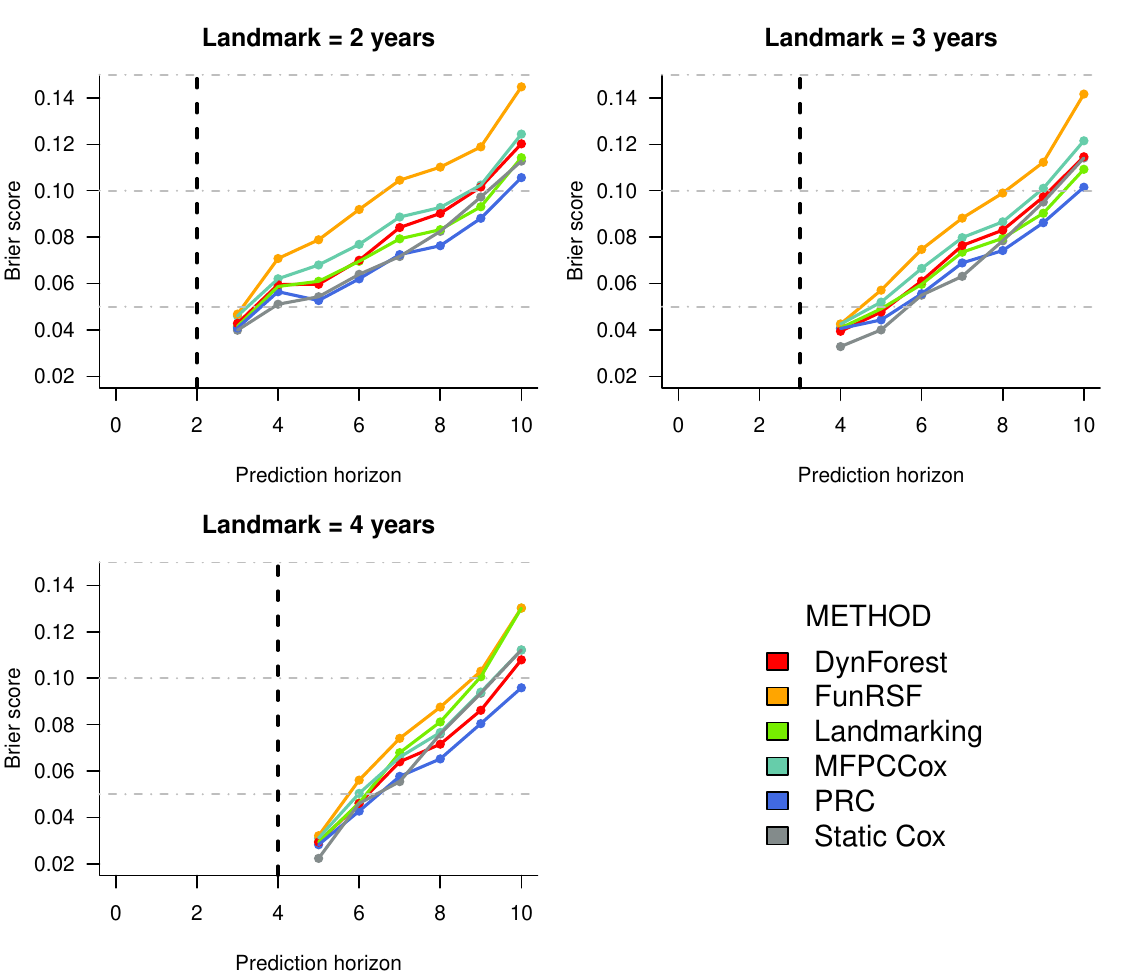}
\end{center}
\caption{Cross-validated Brier score estimates for the prediction of time to dementia in the ADNI dataset.
The corresponding numeric values can be found in Supplementary Table 7.}
\label{ADNI-Brier}
\end{figure}

The RCV estimates of the C index, tdAUC and Brier score for the dynamic prediction of time to dementia in the ADNI study are presented in Table \ref{ADNI-C} and Figures \ref{ADNI-AUC} and \ref{ADNI-Brier}.

\textbf{C index}. As shown in Table \ref{ADNI-C}, for all landmark times PRC consistently achieves the highest performance, and FunRSF the lowest one. The relative performance of the remaining methods varies slightly across landmarks: at landmarks 2 and 3, landmarking and static Cox have the second and third highest C index, and are followed by DynForest and MFPCCox. Instead, at landmark 4 DynForest is the second best method. Overall, the relative performance of DynForest and MFPCCox improves as the landmark time increases,  whereas that of landmarking and static Cox worsens.

\begin{table}[h!]
\centering
\begin{tabular}{lccc}
  \hline
  & \multicolumn{3}{c}{Landmark}\\
Method & 2 & 3 & 4\\  \hline
Static Cox & 0.901 (0.002) & 0.885 (0.006) & 0.856 (0.008) \\ 
  Landmarking & \textbf{0.906} (0.001) & \textbf{0.890} (0.004) & 0.855 (0.009) \\ 
  MFPCCox & 0.889 (0.003) & 0.872 (0.009) & 0.859 (0.008) \\ 
  PRC & \textbf{0.913} (0.001) & \textbf{0.908} (0.003) & \textbf{0.904} (0.003) \\ 
  FunRSF & 0.873 (0.004) & 0.858 (0.011) & 0.845 (0.012) \\ 
  DynForest & 0.891 (0.003) & 0.883 (0.005) & \textbf{0.871} (0.011) \\
   \hline
\end{tabular}
\caption{Cross-validated C index estimates (and standard deviation) for the ADNI dataset. 
For each landmark, the top two methods are highlighted in \textbf{bold}.}
\label{ADNI-C}
\end{table}

\textbf{Time-dependent AUC} (Figure \ref{ADNI-AUC}). 
The results for the tdAUC align closely with the C index. PRC is again the best performing method, and FunRSF the worse, at each landmark time. 
Landmarking is the second best at landmark 2, and DynForest at 4; the two methods have similar performance at landmark 3.  
MFPCCox and FunRSF typically show lower accuracy. 
Once again we can observe that when compared to the multi-step methods, the performance of LOCF landmarking and of the Static Cox model tends to worsen as the landmark increases.

\textbf{Brier score} (Figure \ref{ADNI-Brier}). 
The differences across methods appear to be smaller when looking at the Brier score.
 At all landmarks, the static Cox model performs quite well for predictions until year 7; from year 8, its performance starts to worsen in comparison to other methods. 
 PRC is usually the best or second best performing method, followed by landmarking,  DynForest and MFPCCox. 
 Once again, FunRSF performs worse than all other methods at every landmark time.
 
\textbf{Effect of sample size and number of predictors on predictive performance}.
To investigate the effect of the sample size $n$ on predictive performance, we proceeded to fit the same models on a third and two thirds of the available observations, considering the first landmark (Supplementary Table 12 and Supplementary Figures 4-5). Moreover, we investigated the effect of the number of predictors $P$ and $Q$ on predictive performance by refitting the same models using a third and two thirds of the covariates considered in the full analysis (Supplementary Table 13 and Supplementary Figures 6-7).\\
Overall, the sample size does not affect substantially the cross-validated values of the C index, time-dependent AUC and Brier score, however we can notice that the standard deviation over the repetitions of the CV decreases substantially with $n$; this reflects the fact that as the sample size increases, more information is available for accurate evaluation of the predictive performance.
On the other hand,  increasing the number of predictors leads to higher predictive performance, producing slight increases in the C index and tdAUC, and reductions in the Brier score. This is in line with the expectation that using more predictors can lead to gains in predictive performance (although in this specific dataset the gains appear to be rather small, suggesting that it is already possible to achieve a high predictive performance with a rather small number of predictors). 

\textbf{Summary}. The performance results for the ADNI dataset point out some interesting trends:
\begin{enumerate}
\itemsep0pt
\item landmarking and the static Cox model have a good predictive performance at earlier landmark times (especially at $\ell = 2$). Surprisingly, these methods often outperform MFPCCox, and systematically outperform FunRSF;
\item the two multi-step methods that rely on mixed-effects models (PRC and DynForest) typically outperform those that use MFPCA (MFPCCox and FunRSF);
\item conditionally on the method (LMM, or MFPCA) used to summarize the longitudinal predictors,  the methods that use a Cox model for the survival outcome (PRC and MFPCCox) outperform the corresponding methods that use an RSF (DynForest and FunRSF);
\item overall,  the approach used to model the longitudinal predictors has a stronger impact on predictive performance than the approach used to model the survival outcome.
\end{enumerate}

\subsection{Predictive performance on the ROSMAP data}
\label{res:ROSMAP}

\begin{figure}[h!]
\begin{center}
\includegraphics[scale=0.9]{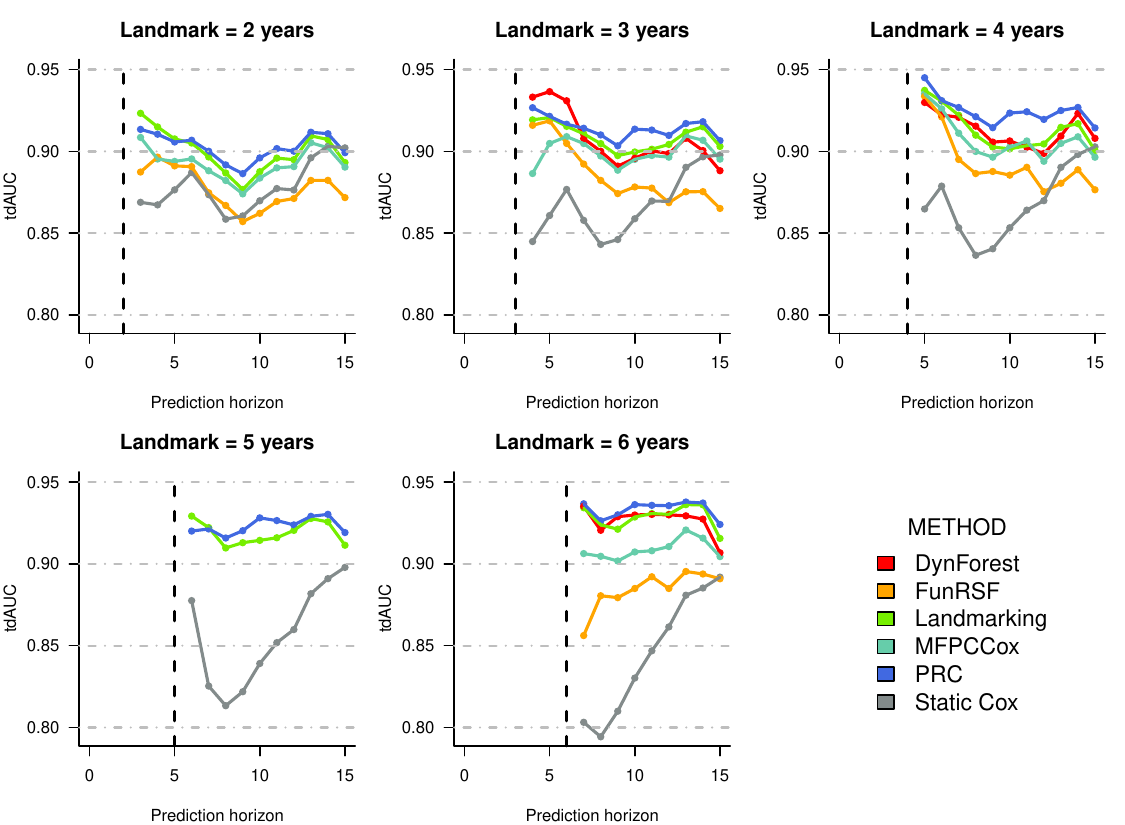}
\end{center}
\caption{Cross-validation estimates of the time-dependent AUC for the prediction of time to AD in the ROSMAP dataset.
The corresponding numeric values can be found in Supplementary Table 8.}
\label{ROSMAP-AUC}
\end{figure}

\begin{figure}[h!]
\begin{center}
\includegraphics[scale=0.9]{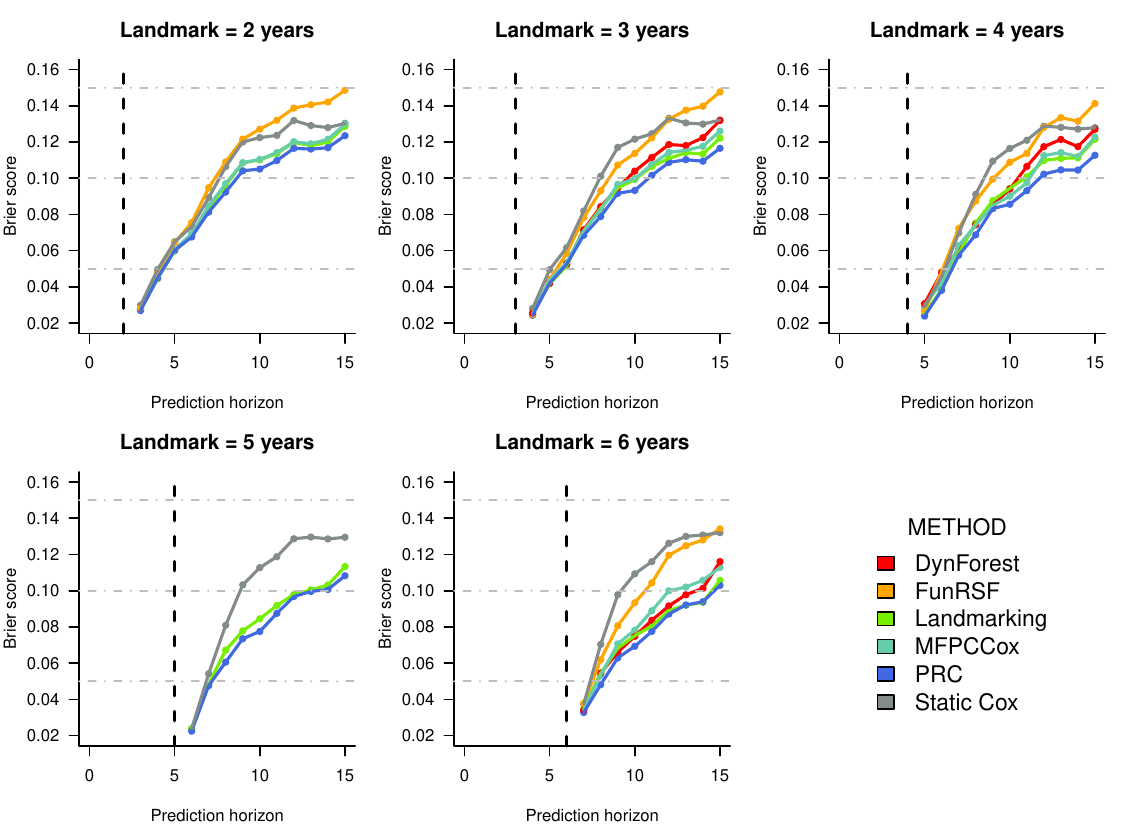}
\end{center}
\caption{Cross-validated Brier score estimates for the prediction of time to AD in the ROSMAP dataset.
The corresponding numeric values can be found in Supplementary Table 9.}
\label{ROSMAP-Brier}
\end{figure}

The RCV estimates of the C index, tdAUC and Brier score for the application to the ROSMAP data are presented in Table \ref{ROSMAP-C} and Figures \ref{ROSMAP-AUC} and \ref{ROSMAP-Brier}.
At landmarks 2 and 5, no results were produced for DynForest due to the fact that estimation of this model failed on all or most (100 and 97 / 100, respectively) folds during the repeated CV. Similarly, at landmark 5 no results are presented for MFPCCox and FunRSF as computation of the MFPCA scores failed on 97 of the 100 folds. These convergence issues are due to a higher percentage of missing visit / values in this dataset than in the ADNI dataset, which complicates the modelling of the longitudinal predictors.

\begin{table}[h!]
\centering
\begin{tabular}{lccccc}
  \hline
  & \multicolumn{5}{c}{Landmark}\\
Method & 2 & 3 & 4 & 5 & 6\\  \hline
Static Cox & 0.824 (0.003) & 0.812 (0.003) & 0.811 (0.003) & 0.799 (0.005) & 0.786 (0.004) \\ 
  Landmarking & \textbf{0.844} (0.002) & \textbf{0.849} (0.002) & \textbf{0.851} (0.001) & \textbf{0.854} (0.001) & \textbf{0.860} (0.001) \\ 
  MFPCCox & 0.837 (0.003) & 0.845 (0.007) & 0.843 (0.027) & - & 0.842 (0.026) \\ 
  PRC & \textbf{0.846} (0.001) & \textbf{0.857} (0.001) & \textbf{0.864} (0.002) & \textbf{0.862} (0.001) & \textbf{0.868} (0.002) \\ 
  FunRSF & 0.806 (0.002) & 0.815 (0.005) & 0.814 (0.024) & - & 0.808 (0.022) \\ 
  DynForest & - & 0.835 (0.027) & 0.836 (0.018) & - & 0.853 (0.003) \\ 
   \hline
\end{tabular}
\caption{Cross-validated C index estimates (and standard deviation) for the ROSMAP dataset.
For each landmark, the top two methods are highlighted in \textbf{bold}.}
\label{ROSMAP-C}
\end{table}

\textbf{C index} (Table \ref{ROSMAP-C}).  
PRC and LOCF landmarking appear to be the best performing methods  at all landmark times. The difference in performance between the two methods is negligible at landmark 2,  and somewhat larger at all other landmarks. 
MFPCCox ranks third at landmarks 2, 3 and 4, and fourth at landmark 6; DynForest ranks fourth at landmarks 3 and 4, and third at landmark 6. FunRSF and the Static Cox model exhibit the worse predictive performance at all landmarks.

\textbf{Time-dependent AUC} (Figure \ref{ROSMAP-AUC}).  
When looking at the tdAUC estimates,  we notice that the performance difference across methods is larger at the later landmark times. 
Differently from the ADNI data, here we can observe that all multi-step methods and landmarking improve over the static Cox model, suggesting that updating prediction based on the longitudinal data may be more beneficial for the ROSMAP data.
We can observe that PRC,  DynForest and LOCF landmarking are the top performing methods across all landmarks, with PRC usually outperforming the other two methods on the later horizon times.
MFPCCox is usually the fourth best performing method, followed by FunRSF and the static Cox model.

\textbf{Brier score} (Figure \ref{ROSMAP-Brier}). 
Similarly to the tdAUC, also for the Brier score estimates  the difference in performance across methods increases with the landmark.  
Moreover, the relative performance of the methods is fairly consistent across landmarks. PRC and LOCF landmarking exhibit the lowest Brier scores, followed by MFPCCox and DynForest.
Once again, FunRSF and the static Cox model have the worst predictive performance

\textbf{Summary}.  
On the ROSMAP data,  incorporating repeated measurements data in the RPM seems to improve predictive performance more than in the ADNI dataset.
The relative performance of the prediction models is somewhat similar to the one observed with the ADNI data, with PRC and LOCF landmarking performing particularly well, followed by DynForest and MFPCCox. Once again, FunRSF appears to be the worst performing method among the multi-step methods considered in this benchmarking.

\clearpage
\newpage
\subsection{Predictive performance on the PBC2 data}
\label{res:PBC2}

\begin{figure}[ht!]
\begin{center}
\includegraphics[scale=0.8]{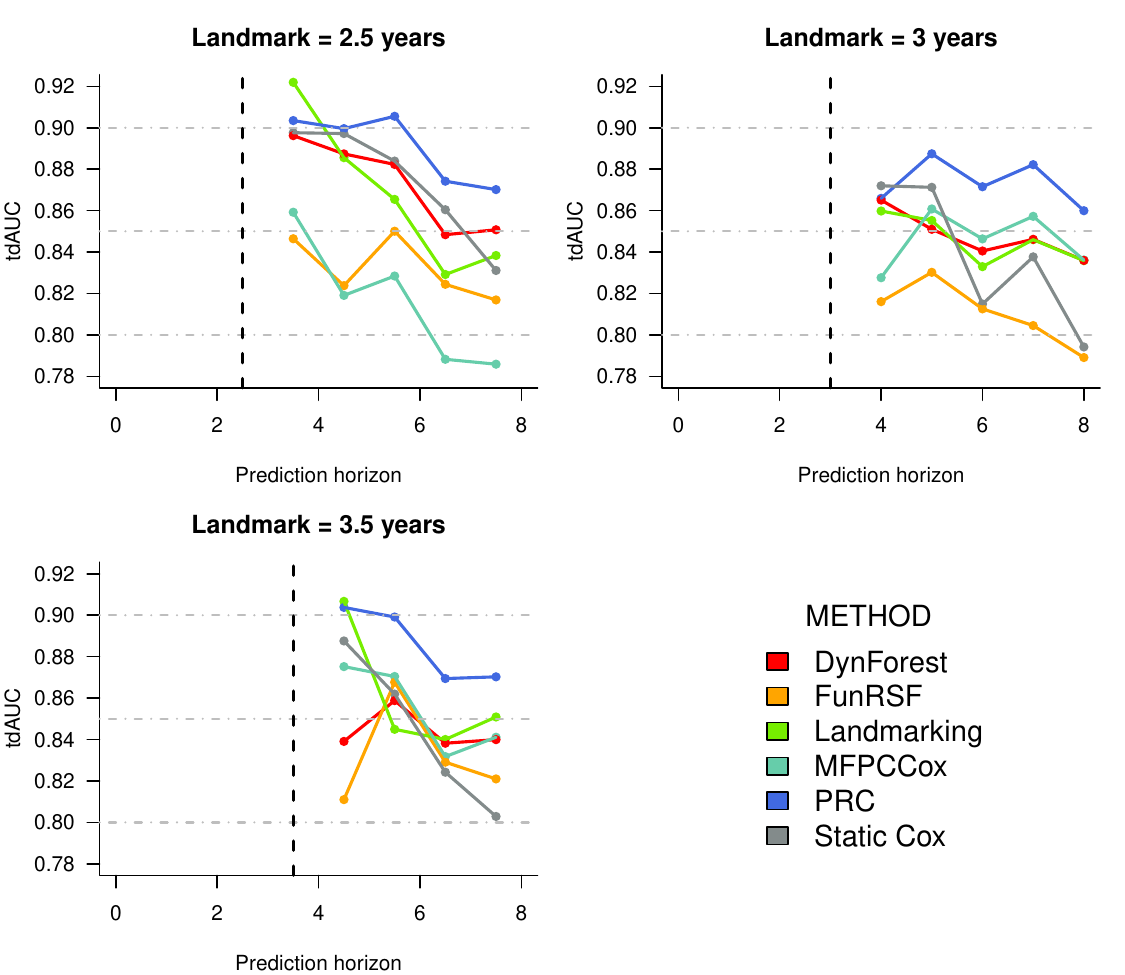}
\end{center}
\caption{Cross-validation estimates of the time-dependent AUC for the prediction of time to death in the PBC2 dataset.
The corresponding numeric values can be found in Supplementary Table 10.}
\label{PBC2-AUC}
\end{figure}

\begin{figure}[ht!]
\begin{center}
\includegraphics[scale=0.8]{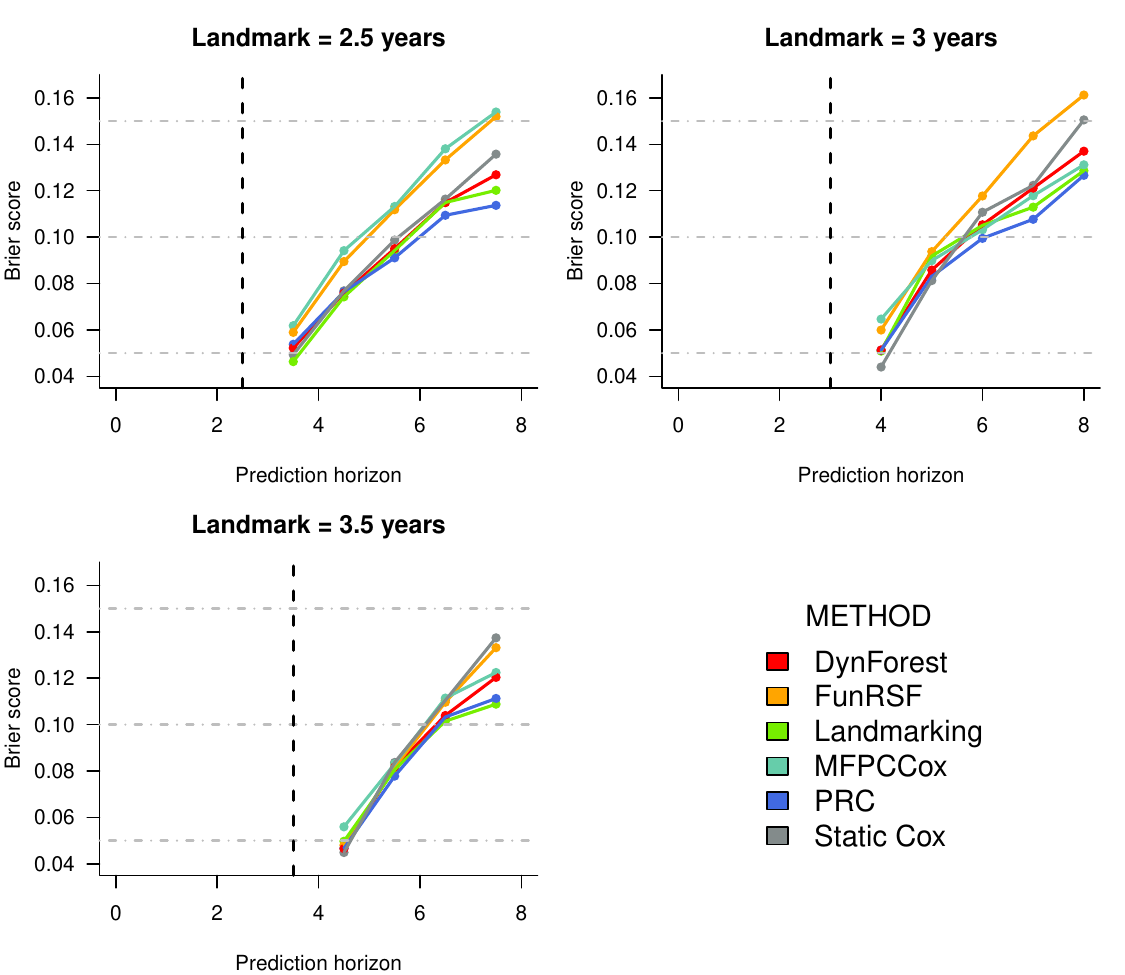}
\end{center}
\caption{Cross-validated Brier score estimates for the prediction of time to death in the PBC2 dataset.
The corresponding numeric values can be found in Supplementary Table 11.}
\label{PBC2-Brier}
\end{figure}

The RCV estimates of the C index, tdAUC and Brier score for the PBC2 dataset are presented in Table \ref{PBC2-C} and Figures \ref{PBC2-AUC} and \ref{PBC2-Brier}.

\begin{table}[ht]
\centering
\begin{tabular}{lccc}
  \hline
  & \multicolumn{3}{c}{Landmark}\\
Method & 2.5 & 3 & 3.5\\  \hline
Static Cox & \textbf{0.813} (0.007) & \textbf{0.791} (0.007) & 0.788 (0.010) \\ 
  Landmarking & 0.798 (0.009) & 0.782 (0.011) & \textbf{0.796} (0.016) \\ 
  MFPCCox & 0.753 (0.022) & 0.786 (0.041) & 0.781 (0.038) \\ 
  PRC & \textbf{0.826} (0.011) & \textbf{0.809} (0.010) & \textbf{0.821} (0.009) \\ 
  FunRSF & 0.760 (0.017) & 0.732 (0.05) & 0.749 (0.035) \\ 
  DynForest & 0.796 (0.012) & 0.766 (0.009) & 0.782 (0.015) \\ 
   \hline
\end{tabular}
\caption{Cross-validated C index estimates (and standard deviation) for the PBC2 dataset.
For each landmark, the top two methods are highlighted in \textbf{bold}.}
\label{PBC2-C}
\end{table}

\textbf{C index} (Table \ref{PBC2-C}).  
At all landmark times, PRC achieves the highest value of the C index. 
Interestingly,  the static Cox model is the second best performing method at landmarks 2.5 and 3 and third best at landmark 3.5, suggesting that the added predictive value of the longitudinal measurements may be more limited for the PBC2 dataset.
LOCF landmarking is the third bist method at landmarks 2.5 and 3, and second best at landmark 3.5.  It is followed by DynForest, MFPCCox and FunRSF. 
By comparing the standard errors in Tables \ref{ADNI-C}, \ref{ROSMAP-C} and \ref{PBC2-C}, we can see that the standard errors around the estimated C index are larger on PBC2 than on ADNI and ROSMAP.  This result can be ascribed to the substantially smaller sample size of this dataset.

\textbf{Time-dependent AUC} (Figure \ref{PBC2-AUC}).  
PRC achieves the highest tdAUC values at all landmarks. 
At landmark 2, it is followed by the Static Cox model, DynForest and LOCF landmarking; FunRSF and MFPCCox have considerably lower predictive accuracy. 
At landmark 3, DynForest, LOCF landmarking and MFPCCox achieve similar tdAUC values.  At landmark 3.5, the trends are somewhat harder to extrapolate. 
Lastly,  at all landmarks the performance of the static Cox model deteriorates quickly as the prediction horizon increases. 

\textbf{Brier score} (Figure \ref{PBC2-Brier}).  
At landmark 2.5, the estimated Brier scores for landmarking, static Cox, PRC and DynForest are very similar, especially at the earlier prediction horizons. The two methods that rely on MFPCA are once again outperformed by all other methods. At landmark 3, the static Cox model has lower Brier score for predictions after 1 and 2 years from the landmark, and PRC for predictions from 3 years from the landmark onwards. At landmark 3.5, the differences between methods tend to be small for horizons at 4.5 and 5.5 years, after which the lowest Brier scores are achieved by landmarking and PRC, and the highest by FunRSF and the static Cox model.

\textbf{Summary}.The estimates of predictive performance for the PBC2 study are generally less accurate than for the ADNI and ROSMAP datasets due to the smaller sample size.  Despite this, the relative performance of the different methods is to a wide extent in line with the patterns observed on the ADNI and ROSMAP datasets.

\subsection{Computing time}
\label{res:time}

We now turn our attention to computing time.
Tables \ref{ADNI-time}, \ref{ROSMAP-time}, and \ref{PBC2-time} report the average computing time required to estimate each model in $k-1$ folds and to obtain predictions for the remaining fold (averaged over all RCV iterations).
Computing time is generally higher for the ROSMAP dataset, where we have the largest sample size and number of predictors, and lowest for the PBC2 dataset where $n$, $P$ and $Q$ are small.
By comparing computing time across landmarks, we observe that the computing time typically decreases as the landmark increases; this is simply an artifact of the fact that at later landmarks less individuals are still at risk, reducing the sample size used for model training and prediction.

Estimation of models that do not model the longitudinal data,  i.e. the static Cox and LOCF landmarking models, is almost immediate: this is because these approaches only require to estimate a Cox model through \texttt{R}'s function \texttt{coxph( )}. 
The multi-step methods are more computationally intensive. Among them,  MFPCCox is the fastest method, followed by FunRSF. This shows that multi-step methods that use MFPCA are faster than those that use LMMs. The faster computing time of MFPCCox is likely due to the use of a Cox model, whose estimation is faster than that of RSF (used by FunRSF).
Lastly, we can observe that among the LMM-based methods,  PRC is substantially faster (between 12 and 20 times, depending on the dataset) than DynForest. This last result is mainly due to the fact that in DynForest, the LMMs are re-estimated at each node split within each random survival tree, making this method considerably slower than all other alternatives.

\begin{table}[h]
\centering
\begin{tabular}{lcccc}
  \hline
  & \multicolumn{3}{c}{Landmark}\\
Method & 2 & 3 & 4 & Average\\  \hline
Static Cox & 0.009 & 0.007 & 0.006 & 0.007 \\ 
  Landmarking & 0.010 & 0.007 & 0.006 & 0.008 \\ 
  MFPCCox & 0.080 & 0.046 & 0.046 & 0.057 \\ 
  PRC & 0.776 & 0.482 & 0.453 & 0.571 \\ 
  FunRSF & 0.240 & 0.122 & 0.125 & 0.163 \\ 
  DynForest & 12.501 & 9.077 & 8.099 & 9.892 \\ 
   \hline
\end{tabular}
\caption{Average computing time per CV fold (in \textbf{minutes}) for the ADNI dataset.}
\label{ADNI-time}
\end{table}

\begin{table}[h]
\centering
\begin{tabular}{lcccccc}
  \hline
  & \multicolumn{5}{c}{Landmark} \\
Method & 2 & 3 & 4 & 5 & 6 & Average\\  \hline
Static Cox & 0.022 & 0.019 & 0.016 & 0.014 & 0.011 & 0.017 \\ 
  Landmarking & 0.023 & 0.020 & 0.017 & 0.014 & 0.011 & 0.017 \\ 
  MFPCCox & 0.194 & 0.135 & 0.066 & - & 0.064 & 0.115 \\ 
  PRC & 3.755 & 1.151 & 1.138 & 1.115 & 1.124 & 1.657 \\ 
  FunRSF & 0.604 & 0.367 & 0.115 & - & 0.124 & 0.302 \\ 
  DynForest & - & 11.504 & 17.657 & - & 34.147 & 21.102 \\ 
   \hline
\end{tabular}
\caption{Average computing time per CV fold (in \textbf{minutes}) for the ROSMAP dataset.}
\label{ROSMAP-time}
\end{table}

\begin{table}[h]
\centering
\begin{tabular}{lcccc}
  \hline
  & \multicolumn{3}{c}{Landmark} &\\
Method & 2.5 & 3 & 3.5 & Average\\  \hline
Static Cox & 0.253 & 0.246 & 0.177 & 0.226 \\ 
  Landmarking & 0.260 & 0.240 & 0.188 & 0.229 \\ 
  MFPCCox & 0.902 & 0.522 & 0.556 & 0.660 \\ 
  PRC & 7.832 & 7.616 & 7.762 & 7.737 \\ 
  FunRSF & 2.776 & 1.393 & 1.526 & 1.898 \\ 
  DynForest & 186.136 & 136.747 & 128.914 & 150.599 \\ 
   \hline
\end{tabular}
\caption{Average computing time per CV fold (in \textbf{seconds}) for the PBC2 dataset.}
\label{PBC2-time}
\end{table}

\newpage
\section{Discussion}
\label{sec5}

This study aimed to benchmark four recently-proposed  multi-step dynamic prediction methods (MFPCCox, PRC, FunRSF, and DynForest) that are capable of using numerous longitudinal predictors to predict survival.  
Each method uses a different combination of MFPCA or LMMs to model the longitudinal predictors, and of Cox models or RSF to predict the survival outcome: therefore, by comparing these methods we can also get a feeling of the extent to which the choices of using MFPCA or LMMs, and a Cox model or an RSF, can affect predictive performance. 
Hereafter we summarize the findings and limitations of this study,  focusing our attention on the modelling flexibility,  predictive performance, and ease of use of the different methods.

\subsection{Modelling flexibility}

The four multi-step methods reviewed in this article represent an important and valuable addition to the dynamic prediction toolbox. 
Differently from traditional landmarking methods, multi-step dynamic prediction methods employ models that enable efficient modelling of the longitudinal data, and to account for the presence of measurement errors. 
These methods are particularly suitable for datasets where where tens, hundreds or (potentially) thousands of longitudinal predictors may be available, where the estimation of JMs is not only very computationally intensive, but usually also unfeasible.

At the same time,  when modelling real-world longitudinal and survival data it may sometimes necessary to deal with competing risks, interval censoring, and missing data. Each of these features can lead to complications with some of the multi-step methods considered in this work, at least in their present state of development:
\begin{enumerate}
\item whereas DynForest has been developed within a competing risks framework,  MFPCCox, FunRSF and PRC don't allow to model competing risks directly. While extending them to competing risks is possible, currently users would need to do that by themselves;
\item none of the four methods currently allows interval censored survival outcomes, which are common in longitudinal studies. Therefore, users need to simplify the survival outcome to right-censored;
\item PRC and DynForest are currently limited to LMMs (and multivariate LMMs in the case of PRC). Extending them to GLMMs would enable a more appropriate modelling of skewed as well as discrete longitudinal covariates, although it can be anticipated that such an extension would be computationally more complex,  potentially leading to more frequent problems during the estimation of the GLMMs and the computation of the predicted random effects;
\item as seen on the ROSMAP dataset, a high percentage of missingness in the longitudinal data can considerably complicate the estimation of MFPCCox, FunRSF and DynForest.  For MFPCCox and FunRSF, this makes it harder to estimate the covariance matrices during the univariate FPCA steps. For DynForest, this creates problems with the estimation of LMMs in children nodes with a small number of subjects.
\end{enumerate}

\subsection{Predictive performance}

The results presented in Section \ref{sec4} are mostly consistent across datasets and performance measures. The most interesting patterns are:
\begin{enumerate}
\item multi-step methods that use LMMs (PRC and DynForest) clearly outperform methods that use MFPCA (MFPCCox and FunRSF) to model the trajectories of the longitudinal covariates.  PRC is frequently the best performing method with respect to the different metrics considered; it is usually followed by DynForest and MFPCCox, while FunRSF almost always exhibits the worst predictive performance;
\item conditionally on the same method being used to summarize the longitudinal covariates, the methods that use a Cox model (PRC and MFPCCox) tend to outperform the corresponding method that uses RSF (DynForest and FunRSF, respectively). However, the difference is substantially smaller than the one between LMM-based and MFPCA-based methods, and it may be dataset specific, as the performance of RSF may improve on datasets with larger sample sizes;
\item interestingly, LOCF landmarking often showed a good predictive performance,  often outperforming some of the multi-step methods. This suggests that this simple and intuitive prediction method can often deliver good predictions of survival. Thus, we recommend that practitioners interested in the use of multi-step methodsir always compare the predictive performance to that of LOCF landmarking, because the latter approach may be preferable in case of similar predictive performance;
\item the extent to which longitudinal data may improve the predictive accuracy of a dynamic RPM over a static RPM varies across datasets. To quantify the extent of this improvement for a specific dataset, we suggest comparing the predictive performance of the developed dynamic RPM to that of a simpler static RPM that only uses baseline information.
\end{enumerate}
 
\subsection{Software and computing time}

Software availability and computing time are two matters of practical importance when attempting to estimate and implement a RPM,  affecting the ease with which a practicioner may be able to estimate a RPM and employ it to compute predictions.

PRC and DynForest come with fairly advanced software implementations and documentation \citep{signorelli2024,devaux2024} that make it easier for users to implement those methods.  On the contrary, MFPCCox and FunRSF lack software implementations and documentation: this forces users to reuse the code presented in the corresponding publications, and often requires them to adapt it to deal with missing data and irregular measurement grids.

MFPCCox and FunRSF are computationally inexpensive: their estimation takes less than a minute on all three datasets considered in our benchmarking. PRC is a bit more intensive, taking less than a minute on PBC2 and ADNI, and approximately 1.66 minutes on ROSMAP. Finally,  DynForest is by far the slowest method: its estimation takes about 2.5 minutes on PBC2, 10 on ADNI, and 21 on ROSMAP.

\subsubsection*{Supplementary material}
\small
Supplementary tables and figures are available in the ``Supplementary Material" pdf file.

\subsubsection*{Code and data}
The \texttt{R} code used to perform the analyses described in this paper is available at\\ \url{https://github.com/mirkosignorelli/comparisonDynamicPred}.
A copy of the PBC2 data is available at the same URL.\\
The data from the ROSMAP study are available upon motivated request from the Rush Alzheimer's Disease Center (RADC) of the RUSH University. ROSMAP resources can be requested at \url{https://www.radc.rush.edu} and \url{www.synpase.org}.\\
The ADNI data are available upon motivated request from the Alzheimer’s Disease Neuroimaging Initiative (ADNI) database (\url{http://adni.loni.usc.edu}). \\

\subsubsection*{Acknowledgements}
\small
\textbf{ROSMAP}.
Data from the ROSMAP study were obtained from the Rush Alzheimer's Disease Center (RADC) of the RUSH University (\url{https://www.radc.rush.edu}). ROSMAP is supported by P30AG10161, P30AG72975, R01AG15819, R01AG17917, U01AG46152, and U01AG61356.  

\textbf{ADNI}.
Data from the ADNI study were obtained from the ADNI database (\url{http://adni.loni.usc.edu}). As such, the investigators within the ADNI contributed to the design and implementation of ADNI and/or provided data but did not participate in analysis or writing of this paper. A complete listing of ADNI investigators can be found at \url{https://adni.loni.usc.edu/wp-content/uploads/how_to_apply/ADNI_Acknowledgement_List.pdf}. 

\bibliographystyle{apalike}
\small
\bibliography{newbiblio}

\end{document}